\documentclass[twocolumn]{openjournal}
\usepackage{natbib}
\usepackage{graphicx}
\usepackage{graphicx,amsmath,amssymb,amstext}
\usepackage{amsbsy,amsfonts,amsthm,color}
\usepackage{xcolor}
\usepackage[colorlinks,linkcolor=blue,citecolor=blue,urlcolor=blue ]{hyperref}
\usepackage{bm}

\newcommand{\mathd}{\ensuremath{\mathrm{d}}}

\newcommand{\calP}{\ensuremath{\mathcal{P}}}

\newcommand{\fatx}{\ensuremath{\boldsymbol{x}}}

\newcommand{\lna}[1]{{#1}}

\begin{document}

% The scaling of weak lensing 2-point function bias with redshift, angular range, and signal-to-noise
\title{The impact of signal-to-noise, redshift, and angular range on the bias of weak lensing 2-point functions}

\author{Amy J. Louca}
%\email{}
%\affiliation{Leiden Observatory, Leiden University, Huygens Laboratory, Niels Bohrweg 2, NL-2333 CA Leiden, The Netherlands.}

\author{Elena Sellentin}
\email{sellentin@strw.leidenuniv.nl}
\affiliation{Leiden Observatory, Leiden University, Huygens Laboratory, Niels Bohrweg 2, NL-2333 CA Leiden, The Netherlands.
}

\begin{abstract}
Weak lensing data follow a naturally skewed distribution, implying the data vector most likely yielded from a survey will systematically fall below its mean. Although this effect is qualitatively known from CMB-analyses, correctly accounting for it in weak lensing is challenging, as a direct transfer of the CMB results is quantitatively incorrect. While a previous study (Sellentin et al. 2018) focused on the magnitude of this bias, we here focus on the frequency of this bias, its scaling with redshift, and its impact on the signal-to-noise of a survey. Filtering weak lensing data with COSEBIs, we show that weak lensing likelihoods are skewed up until $\ell \approx 100$, whereas CMB-likelihoods Gaussianize already at $\ell \approx 20$. While COSEBI-compressed data on KiDS- and DES-like redshift- and angular ranges follow Gaussian distributions, we detect skewness at 6$\sigma$ significance for half of a Euclid- or LSST-like data set, caused by the wider coverage and deeper reach of these surveys. Computing the signal-to-noise ratio per data point, we show that precisely the data points of highest signal-to-noise are the most biased. Over all redshifts, this bias affects at least 10\% of a survey's total signal-to-noise, at high redshifts up to 25\%. The bias is accordingly expected to impact parameter inference. The bias can be handled by developing non-Gaussian likelihoods. Otherwise, it could be reduced by removing the data points of highest signal-to-noise.
\end{abstract}

\maketitle

\section{Introduction}
Weak lensing is a central observable for contemporary cosmology. Arising from the bending of light rays in gravitational fields, it allows not only to study gravity beyond its Newtonian limit, but also to map dark and luminous matter alike. When jointly analyzed with galaxy clustering, it can constrain the gravitational slip parameter \citep{Slip} and thereby discriminate between different types of gravitational theories. Considerable observational efforts are thus invested into weak lensing, including the Kilo Degree Survey \citep{KiDS}, Dark Energy Survey \citep{DES} and the Hyper Suprime-Cam \citep{HSC}.

Weak lensing is sensitive to the cosmological parameter $\sigma_8$, which describes the `clumpiness' of matter in the Universe in the sense that $\sigma_8$ sets the amplitude of the matter power spectrum. Measurements of $\sigma_8$ derived from weak lensing experiments are mildly discrepant with measurements of $\sigma_8$ from the cosmic microwave background (CMB), see \citet{Planck2013,Planck2015,Planck2018}. Since many years, weak lensing yields lower values of $\sigma_8$, and it has been repeatedly proposed that this might be indicative of an unsolved bias in contemporary weak lensing constraints.

The term `bias' usually carries a negative connotation. However, in a previous study \citet{SHH18} had shown that although the summary statistics derived from weak lensing observations are indeed `biased', this bias has here to be understood in the literal sense. An estimator is said to be biased if the value it yields most likely differs from the one it yields on average. An experiment's average outcome then deviates from its most likely outcome. This occurs if the estimator is drawn from a skewed distribution.

\citet{SHH18} had shown that all weak lensing 2-point function estimators will be affected to various degrees by such a literal bias, as they follow a skewed distribution. In consequence, a sound weak lensing data vector will likely display an amplitude that is below average. 

If such weak lensing data were analyzed with the matching skewed likelihood, then parameter inference could proceed self-consistently. However, the current standard weak lensing analysis employs a symmetric Gaussian likelihood instead. When used to analyze a data vector that is biased low, the Gaussian likelihood will force the fit to center on this low amplitude. As $\sigma_8$ scales the amplitude of weak lensing observables, it can thus be expected that a systematically lowered value for $\sigma_8$ is found in the standard analysis.

Quantifying by how much $\sigma_8$ will be biased low -- and removing this bias -- is a highly non-trivial challenge: solving both questions implies the \emph{correct} weak lensing likelihood must be found, rather than a heuristic non-Gaussian substitute.

Accordingly, a skewed baseline likelihood to handle weak lensing observations alongside galaxy clustering and their cross-correlation has been presented in \citet{ManriqueYus1}. It takes the form of a Bayesian Hierarchical Model and at its core lies the same Wishart distribution that also describes the skewed distribution of the low multipoles of the cosmic microwave background (CMB), see \citet{Hamimeche1,Hamimeche2,Upham}. In comparison to CMB data, weak lensing data follow a decidedly more skewed distribution, as was revealed by simulations \citep{Transcovs,SHH18,SLICS} and subsequently by \citet{AnaAndCora}. This additional skewness can be implemented in the skewed baseline likelihood through a free parameter $g_{\rm eff}$, but to date $g_{\rm eff}$ can only be measured from simulations, and computing it from first principles is still unsolved.

In this paper, we thus return to the origin of skewness in weak lensing likelihoods. In particular, we study whether the problematic skewness can be suppressed without losing precision. Although the skewness could be suppressed by invoking the Central Limit Theorem, this is highly suboptimal: inconsiderate averaging will remove information. We therefore opt for the more refined 
approach by filtering, and study whether skewness can be filtered out. Should such a suppression succeed, then developing non-Gaussian likelihoods could be put aside. We here opt for a set of filters known as COSEBIs, as these are constructed to remove further weak lensing systematics. If also able to suppress skewness, a COSEBI-analysis would thus present a highly controllable weak lensing analysis.

The setup of this paper is as follows. In Sect.~\ref{sect:wl} we introduce weak lensing. In Sect.~\ref{sec:filtering} we describe why it can be hoped that filtering may suppress skewness and ensuing biases. In Sect.~\ref{sec:logCosebis} we introduce the essentials of COSEBI filters, and in Sect.~\ref{sec:samples} we create 819 COSEBI samples to estimate their sampling distribution. Our strategy to detect deviations from a Gaussian distribution is described in Sect.~\ref{sec:NG}. From Sect.~\ref{sec:results} we present the main results of this paper: Sect.~\ref{sec:redshift} describes how sampling distribution skewness evolves with redshift, and Sect.~\ref{sec:CMBcomparison} describes the excess skewness of weak lensing measurements in comparison to the CMB case. Sect.~\ref{sec:angularrange} contrasts Euclid- and LSST-like angular ranges with KiDS- and DES-like angular ranges. Sect.~\ref{sec:excess} then studies the excess probability of finding a `low' weak lensing outcome, if a Gaussian likelihood is insisted upon. Our conclusions are presented in Sect.~\ref{sec:conclusions}, and the appendix describes the construction of COSEBIs and answers questions concerning gravity-induced skewness often encountered.

\section{Weak lensing signal and noise}
\label{sect:wl}
Matter between the observer and far-away sources distorts the image of the sources via gravitational lensing. As light from neighbouring sources propagates through similar foreground fields, weak lensing causes the images of neighbouring sources to have a correlated alignment. This correlation can be extracted from observations, see \citet{BartelSchneid,MartinReview}.

Such weak lensing signals carry two types of noise, `shape noise' due to the diversity of galaxy shapes and orientations, and `cosmic variance' due to the stochastic nature of the matter fields transversed by the propagating light.

A weak lensing analysis begins from a \emph{weak lensing galaxy catalogue} that stores for $N$ galaxies their celestial positions $\bm{x}$ together with the two measured ellipticity components $\epsilon_1$ and $\epsilon_2$ and the redshift.
From such a catalogue, the standard analysis of a weak lensing data set computes binned 2-point statistics, such as the correlation functions $\xi_\pm$. If $\theta$ denotes the angular separation on the sky, then the weak lensing correlation functions $\xi_\pm$ can be extracted via the estimator
 \begin{equation}
     \hat{\xi}_{\pm}(\theta) = \frac{\sum_{ij}w_i w_j\left[\epsilon_t(\bm{x}_i)\epsilon_t(\bm{x}_j) \pm \epsilon_{\times}(\bm{x}_i)\epsilon_{\times}(\bm{x}_j) \right]}{\sum_{ij}w_iw_j},
     \label{esti}
 \end{equation}
where $w_{i,j}$ are the weights arising from the shape measurement process on realistic galaxies, \lna{and the subscript $\pm$ is inherited from the plus-minus on the right-hand side}. We use hats to denote noisy quantities.
 
The angular separation $\theta$ and the positions $\bm{x}$ of galaxies in the sky are still two-dimensional only. As our Universe evolves our time, it is advantageous to split a galaxy catalogue radially into subsets of galaxies. This splitting is known as redshift tomography. Denoting the redshift by $z$, the $i$th redshift bin is $n^i(z)$, which is the normalized distribution of the galaxies assigned to the bin $i$.

These redshift bins are then re-expressed as a function of the comoving distance $\chi$,
 \begin{equation}
    g^i(\chi) = \int_{\chi}^{\chi_{\mathrm{h}}} \mathrm{d}\chi' n^i_{\chi}(\chi')\frac{\chi'-\chi}{\chi'},
\end{equation}
known as the weak lensing kernel corresponding to the $i$th galaxy population. 

The light emitted from galaxies per bin will propagate through foreground fields towards the observer. Of these, cosmological theory can predict the matter power spectrum $\mathrm{P}_\delta$. The continuous propagation of light through these foreground fields causes that the weak lensing power spectrum $\mathrm{P}_{\gamma}^{ij}(\ell)$ integrates over the matter power spectrum, weighted by the lensing kernels
 \begin{equation}
     \mathrm{P}_{\gamma}^{ij}(\ell) = \frac{9H_0^4\Omega_m^2}{4c^4} \int_0^{\chi_{\mathrm{h}}} \mathrm{d}\chi \frac{g^i(\chi)g^j(\chi)}{a^2}\mathrm{P}_{\delta}\left(\frac{\ell+1/2}{\chi},\chi\right),
     \label{P}
 \end{equation}
 where the Limber approximation has been used.
Furthermore, $\ell$ is a spherical harmonic multipole, $H_0$ is the Hubble constant, $\Omega_m$ is the matter density parameter, $c$ is the speed of light. Both the matter power spectrum $\mathrm{P}_{\delta}$ and the comoving distance $\chi(z)$ as a function of redshift depend on a broad range of cosmological parameters, which can accordingly be inferred from weak lensing data.

 \section{Filtering to suppress skewness?}
 \label{sec:filtering}
Comparing Eq.~(\ref{esti}) and Eq.~(\ref{P}), an inherent disparity between weak lensing observations and its theoretical predictions becomes apparent: while the weak lensing signal is easiest to extract in real space $\theta$, the theoretical predictions are easier to compute in harmonic space $\ell$. The translation from harmonic space to real space occurs via a filter $F(\ell \theta)$ such that a general weak lensing 2-point function can be written as
\begin{equation}
    \xi^{ij}_{\rm F}(\theta) = \frac{1}{2\pi} \int \mathd \ell \ \ell F(\ell \theta) P_\gamma^{ij}(\ell),
    \label{F}
\end{equation}
where the filter $F$ is to be specified. The correlation function $\xi_+(\theta)$ employs the filter $J_0(\ell \theta)$ and $\xi_-(\theta)$ employs $J_4(\ell \theta)$, which are the zeroth and fourth order Bessel functions respectively. \lna{ In the manner described in \citet{BartelSchneid, MartinReview}, $\xi_\pm$ are the two correlation functions associated to the spin-2 shear field of weak lensing. }

Within the context of determining whether a Gaussian likelihood function is adequate, the filter in Eq.~(\ref{F}) bears a special significance: as shown by \cite{SHH18}, it is the low $\ell$-modes of the transversed matter fields that skew the distribution from which weak lensing data are drawn. Thus, if we succeed in constructing a filter $F(\ell \theta)$ that downweighs the contribution of low $\ell$-modes to a weak lensing observable, then this filter will Gaussianize the distribution.

\begin{figure}
\centering 
\includegraphics[width=0.48\textwidth]{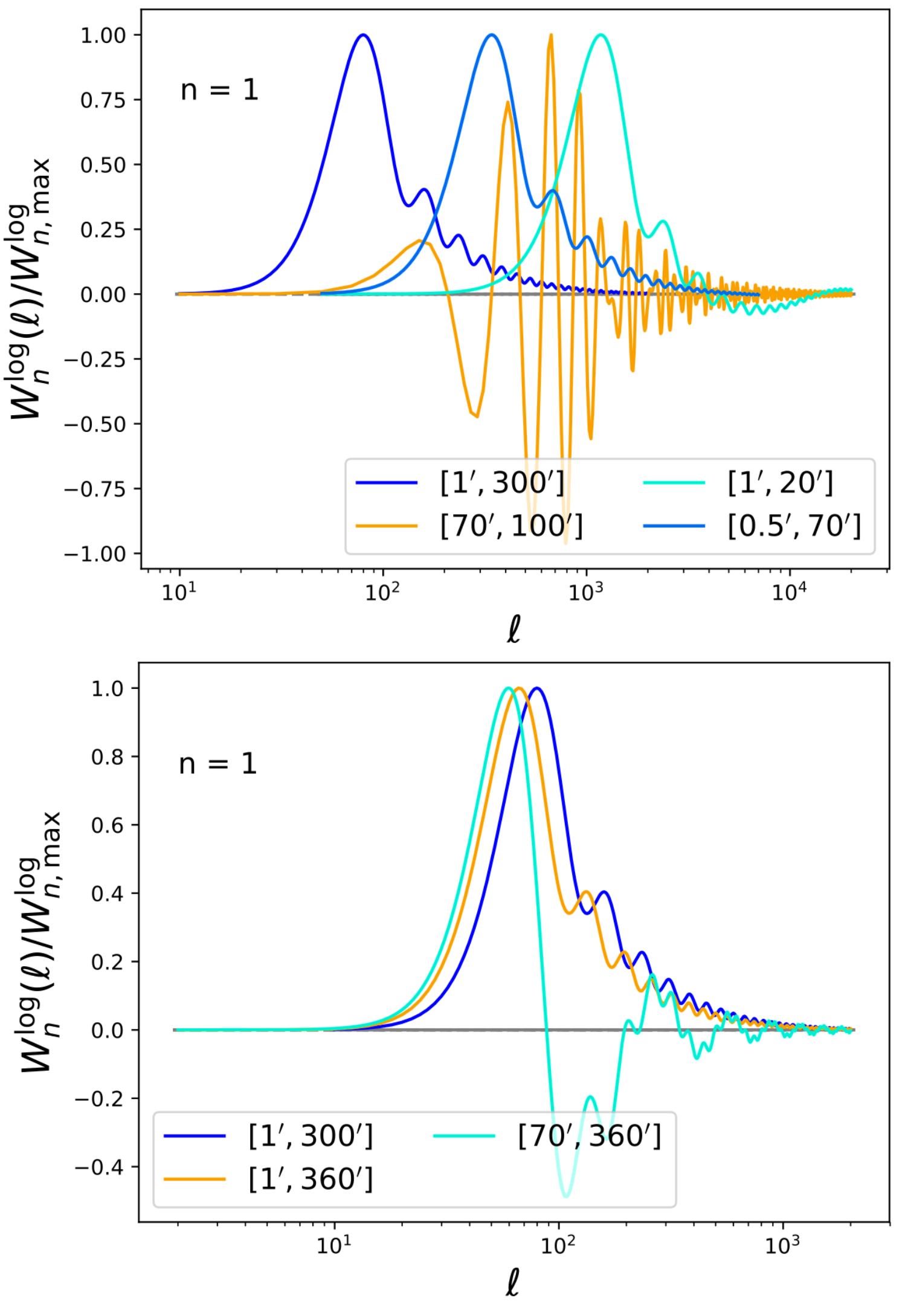}
\caption{The $W^{\rm log}_n(\ell)$ filters for logarithmic COSEBIs, as a function of mode index $n=1$ and multipole $\ell$. The angular range $[\theta_{\rm min}, \theta_{\rm max}]$ is indicated in arc-minutes. It is known that low $\ell$-modes cause non-Gaussianities in 2-point function likelihoods. As low $n$-modes assign large weight to low $\ell$-modes, it can be expected that COSEBIs with low $n$-mode will require non-Gaussian likelihoods. }
\label{Wfilters}
\vspace{10pt}
\end{figure}

The downside to filtering out non-Gaussianity is that if it is done naively (e.g. by setting the filter to zero in the low $\ell$-regime), then constraining power is lost. We thus have to find a filter that is both efficient in retaining cosmological information, \emph{and} in suppressing non-Gaussianity. In this context, the so-called COSEBI filters proposed by \citet{SK07,SEK10} are of interest. Originally developed to split weak lensing E-modes from B-modes, and applied as such to the data of the CFHTLenS \citep{CosebiCFHTLenS}, KiDS and DES survey \citep{CosebiKiDSDES}, COSEBIs are not only efficient in retaining cosmological information, but might as well suppress non-Gaussianity.

\begin{figure*}
\centering 
\includegraphics[width=0.9\textwidth]{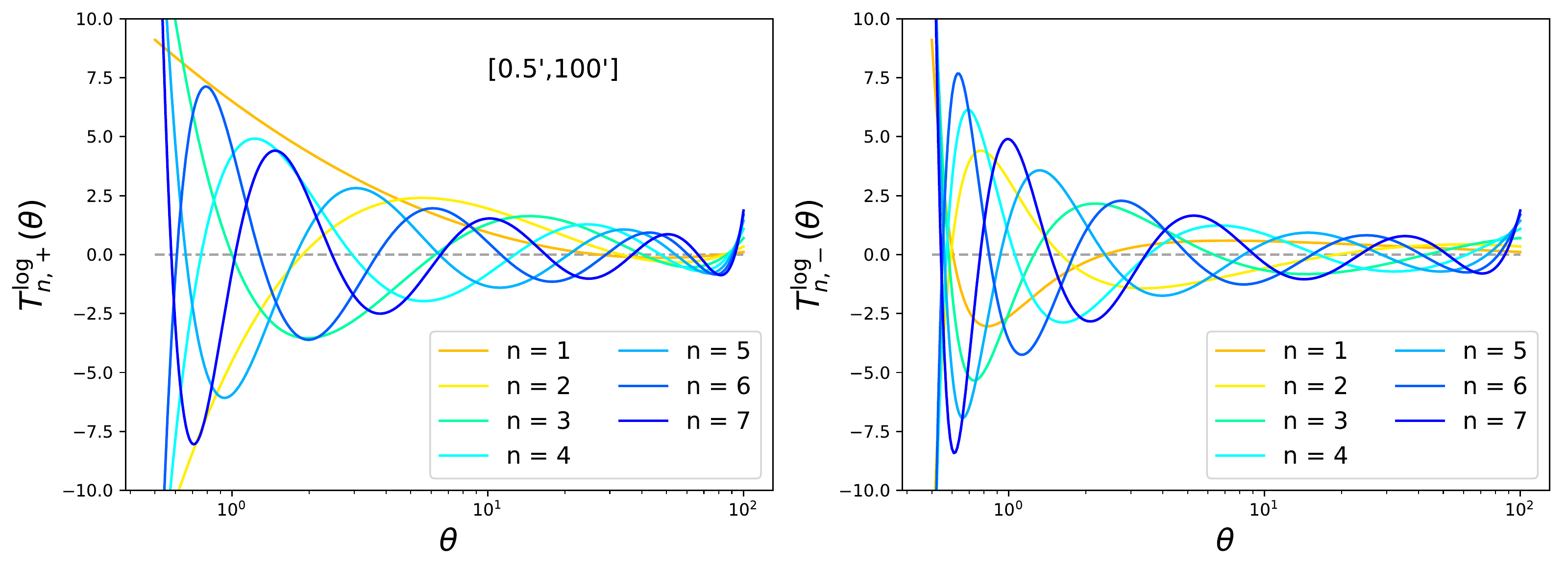}
\caption{The logarithmic filter functions $T^{\mathrm{log}}_{\pm}(\theta)$ as a function of angle for $\theta_{\rm min} = 0.5'$ and $\theta_{\rm max} = 100'$. Filters as these are used to create the first 7 COSEBI modes, $E_n$ and $B_n$, from shear 2-point correlation functions $\xi_{\pm}(\theta)$.}
\label{Tfilters}
\vspace{10pt}
\end{figure*}

~~~~\\
\section{Essentials of logarithmic COSEBIs}
\label{sec:logCosebis}
\textit{Complete Orthogonal Sets of E-/B-mode Integrals}, or COSEBIs for short, were originally introduced to split weak lensing E-modes from B-modes \citep{SK07,SEK10}. The physics of weak lensing is expected to produce E-modes only, and the detection of B-modes would thus directly indicate systematic effects. By discarding B-mode COSEBIs and only analyzing their E-mode counterparts, a clean data vector can be gained and used for parameter inference. However, this clean E-mode data vector still requires the correct likelihood to be known, otherwise the use of the incorrect likelihood becomes a systematic itself.

With the aim of determining the likelihood shape of COSEBIs, we thus study whether the COSEBI filter $W^{\rm log}_n(\ell)$ succeeds in suppressing non-Gaussianity that skews  the likelihood for parameter inference. 

We detail the derivation of COSEBIs in appendix \ref{app:cosebis}. For the study of the COSEBI likelihood shape, it is sufficient to recognize that the COSEBI filters will act according to Eq.~(\ref{F}): given a power spectrum, the filter lets certain $\ell$-modes pass, while others are suppressed, depending on the amplitude that the filter assigns to these modes. 

As it is known that the subclass of logarithmic COSEBIs is more efficient in retaining cosmological information, we here specialize to \emph{logarithmic COSEBIs}, which we denote as $E_n$. COSEBIs are discrete modes, enumerated by their index $n$, and we  expand until $n = 7$, which has been shown to retain the majority of cosmological information \citep{CosebiwCDM}.

In case of logarithmic COSEBIs, the filter $F$ from Eq.~(\ref{F}) is usually referred to as $W_n^{\rm log}(\ell)$. Per index $n$, the filter will output the corresponding COSEBI $E_n$, implying the filter's shape will change when varying $n$. 

As COSEBIs are constructed to only use data on given angular ranges (see appendix \ref{app:cosebis}), the filters $W_n^{\rm log}(\ell)$ implicitly also depend on $\theta_{\rm min}$ and $\theta_{\rm max}$, the minimum and maximum angular range of data that the filter lets pass. 

The exclusion of angular ranges becomes more evident when transforming the $W_n^{\rm log}(\ell)$ filters to real space, where they filter $\xi_\pm(\theta)$ instead of the shear power spectrum $P_\gamma^{ij}(\ell)$. The real space analogues of the $W_n^{\rm log}(\ell)$ filters are here denoted as $T_{+,n}^{\rm log}(\theta)$ and $T_{-,n}^{\rm log}(\theta)$, depending on whether they filter $\xi_+$ or $\xi_-$.

We plot harmonic space representations $W_n^{\rm log}(\ell)$ of selected COSEBI filters in Fig.~\ref{Wfilters}. Fig.~\ref{Tfilters} displays examples of $T_\pm$ filters for $n \in [1,7]$ and an angular range of $0.5$ to $100$ arc-minutes. The angular ranges $[{\theta_{\mathrm{min}}},{\theta_{\mathrm{max}}}]$ that each COSEBI filter integrates over are indicated in the legends of our plots. 

From the harmonic space representation $W_n^{\rm log}(\ell)$ in Fig.~\ref{Wfilters} it is evident that COSEBI filters let a selected range of $\ell$-modes pass. As \citet{SHH18} identified low $\ell$-modes to be the cause of likelihood skewness, COSEBIs might thus be able to exclude these problematic modes. In practice, the filter functions displayed in Fig.~\ref{Tfilters} are applied to correlation functions $\xi_\pm$ estimated from observed galaxy catalogues. In Sect.~\ref{sec:samples}, we will use the $T_\pm$ filters on correlation functions estimated from simulations, with the aim to gain a representative sample of simulated COSEBIs to study the shape of their sampling distribution.

\begin{figure*}
    \centering
    \includegraphics[scale=0.85]{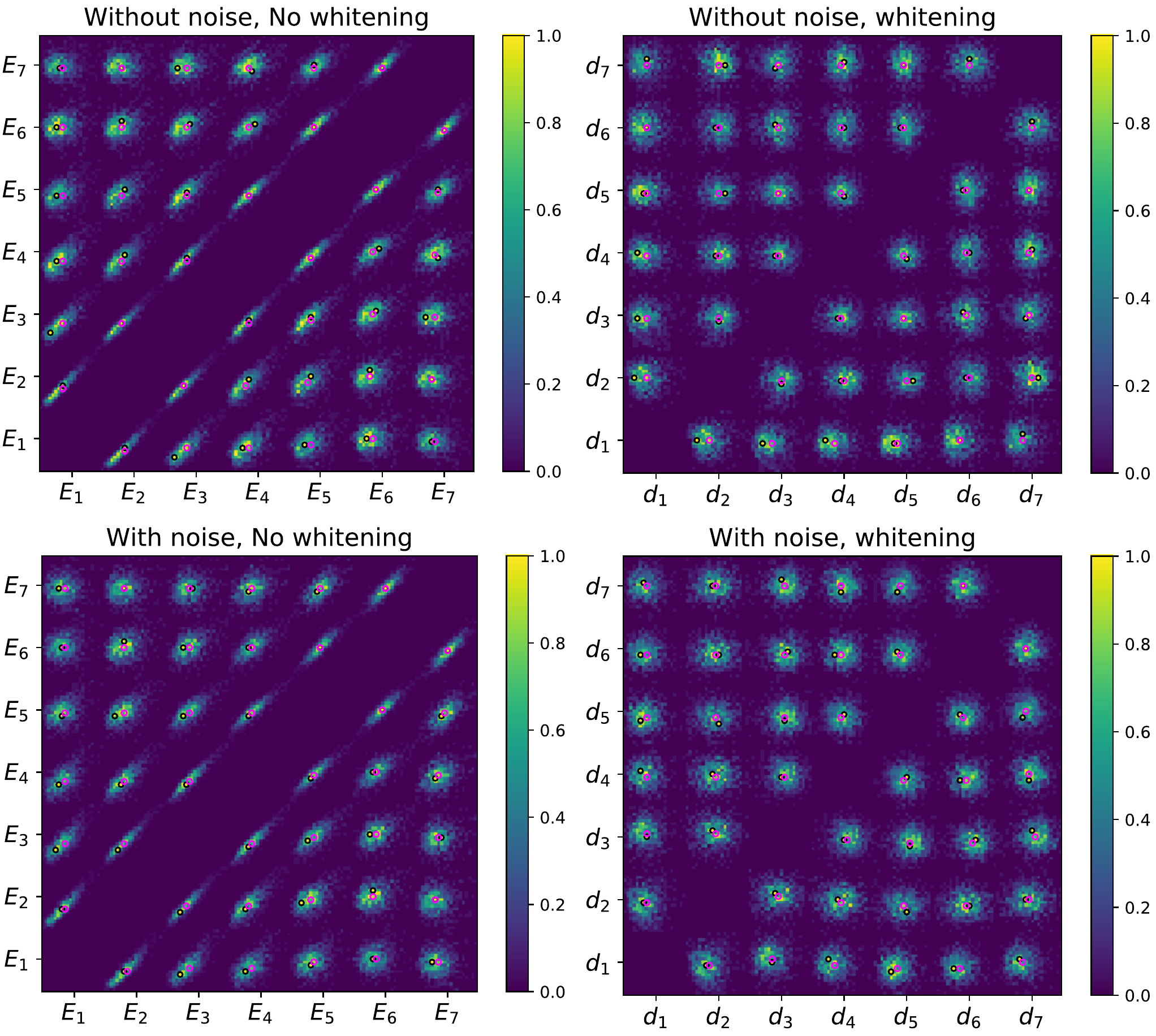}
    \caption{Two-dimensional marginal distributions of the COSEBIs in redshift bin 8. The seven $E_n$ modes filtered the angular scale [1',300']. Per marginal, the two $E_n$ modes are indicated on the axes. The black circles mark the peak of the marginals which are often off-center, due to the asymmetry of the distributions; the pink circles mark the means. When Cholesky-whitening the left column, the right column ensues, where each $d_i$ is a linear superposition of the $E_n$ modes. Cholesky-whitening destroys all Gaussian correlations between data points. This is easily seen close to the diagonal where the strongly ellipsoidal (i.e. correlated) COSEBI distributions were transformed to more isotropic, roundish distributions. Non-Gaussian dependencies between random variables survive Cholesky-whitening, which is why the right column still displays asymmetric non-Gaussian features. These residual non-Gaussianities are quantified in Sect.~\ref{sec:NG}.  }
    \label{fig:Amyplot}
\end{figure*}

\section{Creating COSEBI samples}
\label{sec:samples}
In order to study whether COSEBIs suppress non-Gaussianity in a weak lensing data set, we have to estimate the distribution from which such COSEBIs are drawn. This distribution is called the `sampling distribution' of the data. In parameter inference the sampling distribution reoccurs as the likelihood, when it will be evaluated conditionally on the parameters. 
To estimate the sampling distribution, we compute 819 simulated COSEBI data vectors from the SLICS weak lensing simulations \citep{SLICS} in a Euclid- or LSST-like setup of redshift binning.

The SLICS simulations provide square celestial patches of the cosmic large scale structure of size $A_{\rm sky} = 100~ \mathrm{deg^2}$ at a mass resolution of $2.88 \cdot 10^9 M_\odot/h$. The simulation suit contains 930 independent realizations of a gravitationally evolved dark matter field projected onto lens planes in order to compute the light deflection and therefrom the weak lensing signal per simulation. 

Of these simulations, 819 were post-processed to imitate a 10-bin tomographic weak lensing survey, with $\bar{n} = 2.6/\mathrm{arcmin}^2 $ galaxies per bin. This corresponds to a total of 30 galaxies per square-arcminute as expected for a Euclid-like survey.

To compute COSEBIs, we first run the \textsc{TreeCorr} implementation \citep{TreeCorr} of the $\hat{\xi}_\pm$-estimator given in Eq.~(\ref{esti}) on the SLICS weak lensing catalogues. When computing accurate COSEBIs an unusually high angular resolution is of the essence. Following the recommendation of \cite{CosebiCFHTLenS}, we thus use 10000 linear $\theta$-bins, of width $\Delta \theta = 0.00995'$. We then filter the resulting $\hat{\xi}_\pm(\theta)$ estimates with the $T_\pm^{(n)}(\theta)$ filters according to Eq.~(\ref{EBfromxipm}). This results in 819 estimates of E- and B-mode COSEBIs $E_n$ and $B_n$. We retain the E-modes for further statistical analysis, but discard the B-modes as these would not contain physical information in an analysis of real data.

In our computation of simulated data vectors $\hat{\xi}_\pm$, we observe the known loss of power in $\hat{\xi}_\pm$ as reported in Fig.~6 of \citet{SLICSpowerloss}. This loss of power in turn affects the amplitude to the COSEBIs computed from $\hat{\xi}_\pm$. Fortunately, this is a deterministic rather than a stochastic effect, and is caused by the finite mass resolution and box size of the simulations. As \citet{SLICSpowerloss}, we hence rescale the extracted $\hat{\xi}_\pm$ such that they average to the best-fitting parameters of the original simulation run.

To study shape noise and cosmic variance individually, we estimate pure cosmic-variance $\hat{\xi}_\pm(\theta)$ correlation functions from the SLICS simulations.  As shape noise is understood to be Gaussianly distributed, we process the pure cosmic-variance $\hat{\xi}_\pm$ functions further by adding shape noise directly to $\hat{\xi}_\pm$ rather than per galaxy. 

The shape noise variance per galaxy splits as $\sigma_\epsilon^2 = \sigma_{\epsilon_1}^2 + \sigma_{\epsilon_2}^2$ where the indices $1,2$ indicate the two components of the estimated ellipticities. We use the typical standard deviation per component of $\sigma_{\epsilon_i} = 0.29$. Due to $\hat{\xi}_\pm$ being computed in angular bins, the shape noise variance ${\sf{C}}_{\rm s}$ of $\hat{\xi}_\pm$ is given by
\begin{equation}
    {\sf{C}}_{\rm s}(\theta) = \frac{\sigma_\epsilon^2}{A_{\rm sky} \bar{n}^2 2\pi \theta \Delta \theta},
\end{equation}
where $A_{\rm sky} = 100~ \mathrm{deg}^2$ is the area of the simulated celestial patch and $\bar{n}$ the number density of about 2.6 galaxies per $\mathrm{arcmin}^2$ per tomographic bin.

The correlation functions with shape noise then follow to be
\begin{equation}
    \hat{\xi}^{\rm SN}_\pm(\theta) = \hat{\xi}_\pm(\theta) + \hat{s}(\theta),
    \label{addsn}
\end{equation}
where the random variables $\hat{s}$ are drawn from Gaussian distributions of mean zero and variance $ {\sf{C}}_{\rm s}(\theta)$,
\begin{equation}
    \hat{s}(\theta) \sim \mathcal{G}\left(0,\ {\sf{C}}_{\rm s}(\theta) \right).
    \label{s}
\end{equation}
Note, that Eq.~(\ref{addsn}) and Eq.~(\ref{s}) imply that shape noise does not change the signal, but only enhances the scatter around the signal: the quantity $\hat{s}(\theta)$ is a Gaussian random variable, taking with equal probability values above and below its mean.

Both the correlation functions with and without shape noise are then filtered according to Eq.~(\ref{EBfromxipm}) in order to yield 819 E-mode COSEBIs for each $n$ and each redshift bin combination. With 10 redshift bins, there will be 55 redshift bin combinations. As we compute 7 COSEBIs per redshift bin combination, we thus create 819 samples of a 385 dimensional data vector.

As an example, Fig.~\ref{fig:Amyplot} depicts all two-dimensional marginal distributions of the first seven COSEBI E-modes of the redshift bin 8 correlated with itself. This figure can be repeated for the remaining 54 redshift bin combinations, which provides the basis for the non-Gaussianity detection in Sect.~\ref{sec:NG}. 

From Fig.~\ref{fig:Amyplot}, we can already deduce by eye that the assumption of a Gaussian distribution for COSEBIs there used will be approximate. This is evidenced by the marginals deviating from two-dimensional Gaussian distributions. It can be seen that the distributions' peaks are often off-center, which affects in particular the COSEBIs $E_2$ and $E_3$. Additionally, the marginals including the COSEBI $E_6$ display a sharp drop in intensity, causing an edge in the depicted histograms. These first visual impressions of non-Gaussianity will be quantified in Sect.~\ref{sec:NG}.

\section{Non-Gaussianity detection strategy}
\label{sec:NG}
The purpose of this section is to quantify the level of non-Gaussianity in the high-dimensional data set of seven COSEBIs per 55 redshift bin combinations of a Euclid- or LSST-like survey. For detected non-Gaussianities, a follow-up study of the ensuing biases is presented in Sect.~\ref{sec:excess}.

A covariance matrix is sufficient if there \lna{exist only Gaussian correlations in pairs of data points.} The components of the covariance matrix are estimated via the usual sample estimator
\begin{equation}
{C}_{mn}= \frac{1}{N-1}\sum_{i=1}^N (E_m^i -\bar{E}_m )(E_n^i -\bar{E}_n ),
\label{covi}
\end{equation}
which uses the sample mean
\begin{equation}
\bar{E}_n  = \frac{1}{N}\sum_{i=1}^N E_n^i.
\end{equation}
If the likelihood truly is Gaussian, then computing an accurate covariance matrix is the final stepping stone for an unbiased analysis. In contrast, if the likelihood is non-Gaussian, then even an arbitrarily precise covariance matrix is insufficient to parameterize the actual likelihood. The purpose of a \emph{trans-covariance matrix}, as developed by \citet{Transcovs}, is to flag where in a data set computing a covariance matrix is insufficient. If the likelihood truly is Gaussian, then the trans-covariance matrix will vanish. Otherwise, if non-Gaussian dependencies affect the data, then the trans-covariance matrix will flag pairs of data points for which more than their covariance must be understood to yield the correct likelihood.

The logic of a trans-covariance matrix is to first destroy the Gaussian correlations in pairs of data points, and then measure whether any further non-Gaussian dependencies between the two data points remain. If $\boldsymbol{d} = (d_a, d_b)$ is a pair of two data points taken from a usually much larger data vector $\fatx$, then the covariance matrix of this pair is given by
\begin{equation}
   { \sf{C}} = 
   \begin{pmatrix}
   {\rm var}(d_a) & {\rm cov}(d_a,d_b) \\
   {\rm cov}(d_a,d_b) & {\rm var}(d_b)
   \end{pmatrix}.
   \label{cov}
 \end{equation}
This covariance matrix is Cholesky decomposed as $ { \sf{C}} = {\sf L L}^T$, and the pair $\boldsymbol{d}$ is whitened via
\begin{equation}
    \tilde{\boldsymbol{d}} = {\sf L}^T \boldsymbol{d}.
\end{equation}
\lna{If the covariance matrix of Eq.~(\ref{cov}) fully captures the statistical dependence between $d_a$ and $d_b$, then the whitened vector $ \tilde{\boldsymbol{d}}$
will now consist of two independent normally distributed random variables.} This is testable. If the test is failed, non-Gaussian dependencies between the pair exist.

The test for non-Gaussian dependencies must be repeated for every pair of data points. This repetition builds up a matrix: just like the covariance matrix computes the covariance between a pair, the trans-covariance matrix flags the non-Gaussian dependencies in a pair. Plotting the two matrices side by side directly reveals for which pairs the covariance matrix is an all-encompassing statistical description, and which pairs require an advanced statistical treatment.

\begin{figure*}
    \centering
    \includegraphics[scale=0.54]{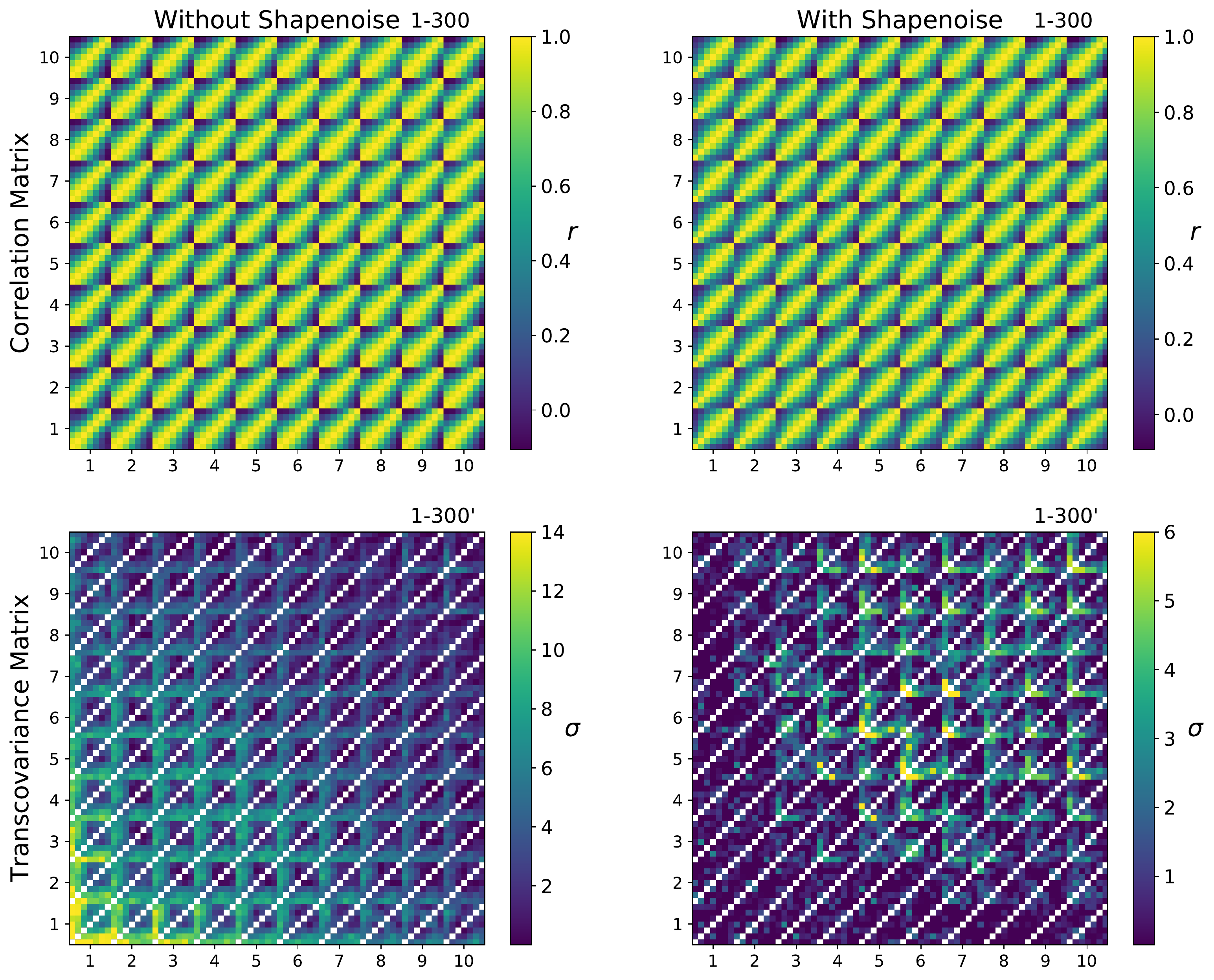}
    \caption{\textbf{Description:} The correlation- and trans-covariance matrices of the seven $E_n$ COSEBIs for all 55 redshift bin combinations of a 10-bin survey. These COSEBIs were computed by filtering $\hat{\xi}_\pm(\theta)$ binned in 250 angular bins. The axis labels indicate the redshift bins; within each redshift bin the mode index $n$ increases from 1 (left) to 7 (right). The left column singles out cosmic variance; the right column also adds shape noise. The angular ranges $[\theta_{\rm min},\theta_{\rm max}]$ of the COSEBIs are indicated. \textbf{Interpretation:} While all COSEBIs are strongly correlated as seen in the upper row, only selected COSEBIs show additional non-Gaussian dependencies. Without shape noise, non-Gaussianity is detected in the low redshift bins with up to 21$\sigma$ significance. For visibility, the colour bar is cut at 14$\sigma$. With shape noise, a redshift trend for non-Gaussianity emerges: While the redshift bins 1-4 are unaffected, higher bins show a $6\sigma$ detection of non-Gaussianity. The redshift bins 1-4 all fall below a redshift of unity, bin 5 marks the transition, and from bin 6 onwards the binned galaxies reside at redshifts $z>1$. \lna{The exact redshift bin boundaries here used are inherited from the SLICS simulations and are stated in Fig.~A1 of \citep{SLICSpowerloss}. They imitate the standard 10 bin setup of a Euclid-like survey.} While the bulk of galaxies of contemporary surveys is located below a redshift of unity, half of the galaxies in a Euclid- or LSST-like survey will reside at redshifts beyond unity. This implies the next generation of surveys will be affected by the found skewness.}
    \label{fig:CorrTransLowRes}
    \vspace{10pt}
\end{figure*}

Our test for non-Gaussian dependencies in a pair proceeds as follows. Assume $\tilde{d}_a, \tilde{d}_b$ indeed were whitened normal distributed variables. Then
\begin{equation}
    \tilde{d}_a \sim \mathcal{G}(0,1), \ \  \tilde{d}_b \sim \mathcal{G}(0,1),
    \label{GG}
\end{equation}
would hold. A direct consequence of this assumption is that the sum of $\tilde{d}_a, \tilde{d}_b$ follows
\begin{equation}
\tilde{d}_a +  \tilde{d}_b \sim \mathcal{G}(0,2).
\label{DD}
\end{equation}
If there exist non-Gaussian dependencies between $d_a$ and $d_b$, then the assumption that $\tilde{d}_a, \tilde{d}_b$ are independently drawn from the two normal distributions in Eq.~(\ref{GG}) breaks down, and Eq.~(\ref{DD}) will be incorrect as a consequence. From our simulated COSEBIs, we can create a histogram of the sum in Eq.~(\ref{DD}). If this histogram tends to a Gaussian with variance two, then the pair $d_a$ and $d_b$ is fully described by its covariance matrix. If the histogram deviates from a Gaussian of variance two, then $d_a$ and $d_b$ depend on each other in ways not captured by a Gaussian likelihood.

Denoting the histogram as $\mathcal{H}(\tilde{d}_a + \tilde{d}_b)$, its deviance from $\mathcal{G}(0,2)$ is computed by integrating over the difference between the histogram and the Gaussian. If the histogram consists of $W$ rectangular bins, then this integral is simply given by the sum
\begin{equation}
  S_{ab} = \frac{1}{W} \sum_{w=1}^W \left[ \mathcal{H}_w\left( \tilde{d}_a +  \tilde{d}_b \right) - \mathcal{G}\left(\tilde{d}_a +  \tilde{d}_b|0,2\right) \right]^2,
\end{equation}
where $\mathcal{H}_w$ is the $w$th histogram bin. The sum $S_{ab}$ provides the $(a,b)$ component of the trans-covariance matrix. The corresponding components of the covariance matrix are given in Eq.~(\ref{covi}).

Finally, we have to mitigate the impact of having only 819 simulations, which will cause noise in the trans-covariance matrices. We thus compute trans-covariance matrices for 819 Gaussian data sets, and measure the mean and standard deviation of their matrix elements. We then mean-subtract the trans-covariance matrices of the actual COSEBI simulations, and divide by the standard deviation. This yields a trans-covariance matrix in units of a standard deviation, such that we can speak of a `$5\sigma$' detection, etc. The units $\sigma$ are thus assigned to the colour bars of all plotted trans-covariance matrices.

\section{Results: Biases and skewness in COSEBIs}
\label{sec:results}
In this section we present the main findings of this paper, which include the scaling of biases from skewness with redshift, $\ell$-mode, and angular range $\theta_{\rm min}, \theta_{\rm max}$ in a weak lensing data set.

\begin{figure*}
    \centering
    \includegraphics[scale=0.8]{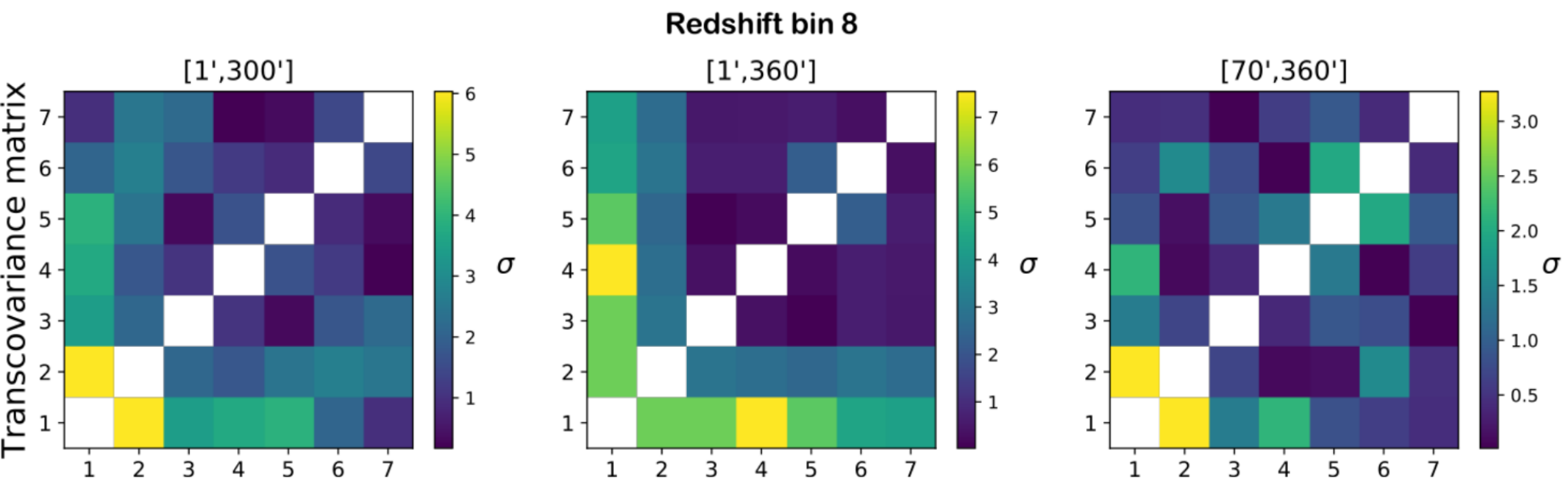}
    \caption{\textbf{Description:} Three magnifications of redshift bin combination 8-8, recomputed from $\hat{\xi}_\pm(\theta)$ with 10000 $\theta$-bins to show that the results with 250 $\theta$-bins are accurate. The axis labels enumerate the seven COSEBI modes. The angular ranges $[\theta_{\rm min},\theta_{\rm max}]$ of the COSEBIs are indicated per panel. \textbf{Interpretation:}  The strength of non-Gaussianity indeed depends on the $\ell$-range the W-filter assigns large weight. Comparing to Fig.~\ref{Wfilters}, the [1',300']-filter amplifies $\ell$-modes around $\ell \approx 100$ for which non-Gaussian sampling distributions are identified with 6$\sigma$ significance (left). The [1',360']-filter emphasizes slightly lower $\ell$-modes causing a non-Gaussianity detection significance of  more than $7\sigma$ (middle). The [70',100']-filter changes sign, thereby balancing positive and negative skewness against each other; this results in a suppression of non-Gaussianity that is consequently only detected at slightly more than 3$\sigma$ (right).}
    \label{fig:3transcovs}
    \vspace{10pt}
\end{figure*}

\subsection{Increase of skewness with redshift}
\label{sec:redshift}
In Fig.~\ref{fig:CorrTransLowRes} we depict the correlation matrices and the trans-covariance matrices for a Euclid- or LSST-like survey with 10 tomographic redshift bins. The redshift bins 1-4 of this simulated survey fall below a redshift of unity. The redshift bin 5 marks the transition to a redshift of unity, and the redshift bins 6-10 finally include galaxies above a unit redshift. For a contemporary survey such as KiDS and DES, the vast majority of their observed galaxies used for weak lensing fall below a unit redshift, see e.g. \cite{Edo, KiDS, DES}.  Accordingly, these surveys can roughly be recognized in the first four redshift bins of Fig.~\ref{fig:CorrTransLowRes}.

Fig.~\ref{fig:CorrTransLowRes} displays the COSEBI modes $E_1$ to $E_7$ per redshift bin combination. The angular range of these COSEBIs is $\theta_{\rm min} = 1'$ and $\theta_{\rm max} = 300'$.

From the correlation matrices with and without shape noise we recognize the familiar trend that weak lensing observations in different redshift bins are strongly correlated. The trans-covariance matrix without shape noise (lower left) reveals a  14$\sigma$ to 21$\sigma$ detection of a non-Gaussian sampling distribution for COSEBIs at low redshift. In the case at hand, this non-Gaussianity takes the form of a skewed sampling distribution, as is expected for any 2-point function, see \citet{SHH18} and references therein. This shape-noise-free panel is included to show the upper bound on non-Gaussianity, as it would occur in a survey of many more galaxies than detectable by Euclid or LSST.

The lower right trans-covariance matrix adds shape noise for a Euclid- or LSST-like survey. Shape noise follows a Gaussian distribution and convolves the originally strongly skewed distributions that gave rise to the 14 to 21$\sigma$ detection on the left. The convolution results in a suppression of non-Gaussianity, and accordingly the trans-covariance matrix with shape noise flags a 6$\sigma$ detection of residual skewness, which is still considerable. All pairs of data points that are flagged as subject to non-Gaussian dependencies will require a non-Gaussian likelihood for bias-free inference.

The perhaps most important result of Fig.~\ref{fig:CorrTransLowRes} is the evidence that a COSEBI-based KiDS and DES analysis on the angular scale $[1',300']$ is free from non-Gaussian biases. This conclusion is evidenced by the first four redshift bins not displaying any prominent non-Gaussianity detection. In contrast, for a Euclid- or LSST-like survey, half of the galaxies reside at redshifts above unity, and will thus be affected by the skewness of sampling distribution and likelihood here detected.

The computation of Fig.~\ref{fig:CorrTransLowRes} was only feasible by binning the $\hat{\xi}_\pm(\theta)$ functions in 250 angular bins before applying the COSEBI filters $T_\pm(\theta)$. While 250 angular bins in $\hat{\xi}_\pm$ are insufficient to yield a percent-level accuracy of the \emph{mean} COSEBI signal (see \citet{CosebiCFHTLenS}), we find that the COSEBI trans-covariance matrix converges much more rapidly than the mean signal. To evidence that 250 angular bins were sufficient, we display in Fig.~\ref{fig:3transcovs} a recomputation of the trans-covariance matrix of redshift bin 8, but now with 10000 angular bins. This yields precisely the same detection significance of the found non-Gaussianities. Further recomputations of the first three redshift bins at angular resolution of 10000 angular bins are presented in Sect.~\ref{sec:angularrange}.

\begin{figure*}
    \centering
    \includegraphics[scale=0.9]{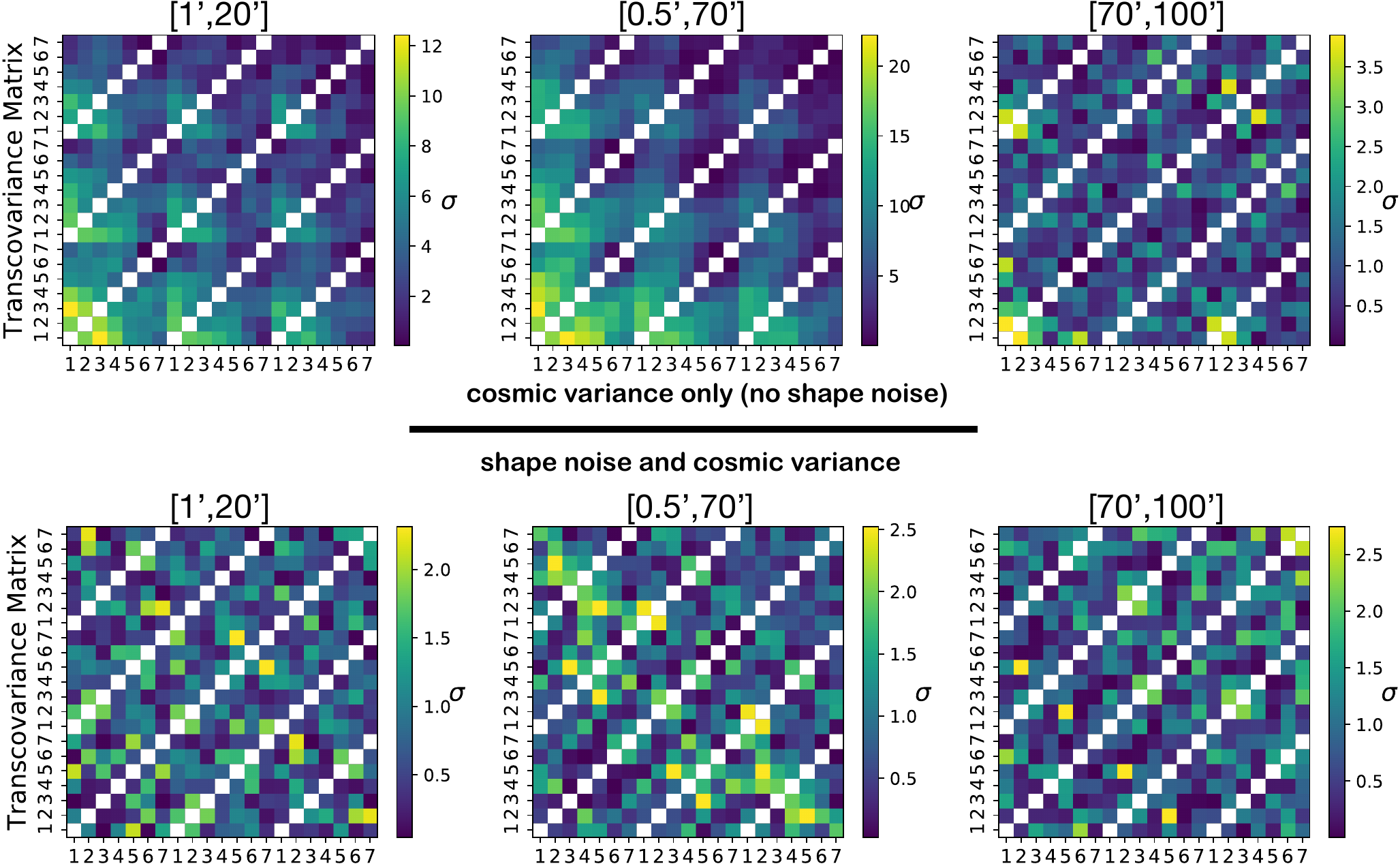}
    \caption{The trans-covariance matrices of the COSEBI E-modes on an angular scale of 0.5' - 20' (left), 0.5'-70' (middle), 70'-100' (right) for redshift bins 1 to 3 (and their cross-correlations) of a Euclid- or LSST-like survey. These three redshift bins extend until $z\approx 1$ and cover the bulk of KiDS- and DES-like galaxy catalogues. The top row depicts E-modes without shape noise, the lower row includes it.  The comparison reveals that although the KiDS- and DES-like redshift ranges are affected by non-Gaussianity from cosmic structure formation, it is the addition of shape noise that renders this skewness negligible. This is evidenced by the up to $18\sigma$ detections of skewness without shape noise (middle top panel), which is suppressed as soon as shape noise is added. Accordingly, the lower row shows no significant detection of skewness throughout the depicted redshift ranges. }
    \label{fig:diffang}
    \vspace{10pt}
\end{figure*}

\subsection{$\ell$-ranges: Excess skewness in comparison to the CMB}
\label{sec:CMBcomparison}
%Discuss 3-panels at 10000 theta bins
% Excess skewness to CMB case
% Consistent with ell-mode cases before
The sampling distribution (and in consequence the likelihood) of any 2-point function is a priori expected to be skewed. This arises due to 2-point functions being quadratic in the underlying field values that they correlate, and in consequence even affects the power spectrum estimates $C_\ell$ of the cosmic microwave background (CMB), although the microwave background itself is compatible with a Gaussian random field \citep{Hamimeche1, Hamimeche2}.

Known from analyses of the CMB is that the sampling distribution of $C_\ell$ is skewed until $\ell \approx 20$, see e.g. the discussion in the Planck likelihood of \citet{PlanckLikelihood2018}. As COSEBI filters allow us to single out certain $\ell$-ranges with more precision than could be achieved with a $\xi_\pm(\theta)$ analysis, we are able to compare likelihood skewness in weak lensing surveys to likelihood skewness in CMB analysis. This study is depicted in Fig.~\ref{fig:3transcovs}.

All three panels of Fig.~\ref{fig:3transcovs} depict the trans-covariance matrix of the COSEBI modes $E_1$ to $E_7$ for redshift bin 8. These COSEBIs were computed by filtering $\xi_\pm$ binned in 10000 angular bins, as described in appendix \ref{app:cosebis}. From left to right, the angular range of the COSEBIs was varied, to the effect of singling out different $\ell$-ranges. The left trans-covariance matrix reveals a 6$\sigma$ detection of skewness on the angular range $[1',300']$. The corresponding $W^{\rm log}_n(\ell)$ filter has been displayed in blue in Fig.~\ref{Wfilters}. Evidently, this filter emphasizes $\ell$-modes in the range $\ell \approx 80-150$, and continues to assign significant amplitude to $\ell$-modes at $\ell \approx 200$. Weak lensing is therefore subject to likelihood skewness over an $\ell$-range roughly ten times larger than for the CMB.

The middle panel of Fig.~\ref{fig:3transcovs} depicts the trans-covariance matrix for $\theta_{\rm min} = 1'$ and $\theta_{\rm max} = 360'$. The corresponding $W^{\rm log}_n(\ell)$ filter is depicted in yellow in the lower panel of Fig.~\ref{Wfilters}. This filter emphasizes $\ell$-modes slightly below those of the COSEBIs in the left panel. Accordingly, non-Gaussianities are now detected with more than $7\sigma$ significance, rather than $6\sigma$ as in the case of the previously discussed $[1',300']$ COSEBIs.

The right-most trans-covariance matrix of Fig.~\ref{fig:3transcovs} employs the angular range $[70',360']$, whose $W^{\rm log}_n(\ell)$ filter is depicted in green in Fig.~\ref{Wfilters}. This filter also assigns a large amplitude on $\ell$-modes around $\ell \approx 100$, but additionally shows a reversal of sign. As neighbouring ${\rm P}_\gamma(\ell)$-estimates are highly correlated in weak lensing, this filter will thus sum over nearly identical ${\rm P}_\gamma (\ell)$ values which it weights with positive and negative signs. As the skewness of a distribution is reversed when the sign of the random variable is reversed, this leads to a cancellation of the skewness during the summation. Accordingly, the right-most trans-covariance matrix of Fig.~\ref{fig:3transcovs} only displays a $3.2\sigma$ detection of residual skewness.

\subsection{Scaling with angular range}
\label{sec:angularrange}
% Angular range implicit in COSEBIS, T-filter do not let points pass outside range; refer to figs at end of paper
Euclid and LSST will not only compile galaxy catalogues that reach deeper into the cosmic past than those of KiDS and DES, but will additionally cover a larger celestial area. Accordingly, it is natural that KiDS and DES limit their analysis of weak lensing correlation functions to smaller angular ranges than we discussed in Sect.~\ref{sec:redshift}.

In Fig.~\ref{fig:diffang} we therefore depict trans-covariance matrices for KiDS- and DES-like redshift ranges and angular ranges. The depicted COSEBIs are $E_1$ to $E_7$, computed for the first three redshift bins of Fig.~\ref{fig:CorrTransLowRes}. The upper limit of these redshift bins is $z \approx 1$. We depict three sets of COSEBIs, with those of the left column spanning the angular range $[1',20']$, the middle column using $[0.5',70']$ and the right-most column $[70',100']$. All these COSEBIs were computed from $\hat{\xi}_\pm(\theta)$ binned in 10000 angular bins.

The upper row of Fig.~\ref{fig:diffang} reveals that these low redshifts are subject to strong non-Gaussianity when cosmic variance alone is regarded. The skewness of the sampling distribution of COSEBIs with $\theta_{\rm min} = 1'$ and $\theta_{\rm max} = 20'$ is detected with more than 12$\sigma$ significance. For the angular range $[0.5',70']$ the skewness of the COSEBI distribution is even detected with 21$\sigma$ significance. On angular ranges $[70',100']$ small clusters of $3.5\sigma$ significance appear, but otherwise the sampling distribution of these COSEBIs is compatible with a Gaussian.
    
The lower row of Fig.~\ref{fig:diffang} reveals that the addition of shape noise suppresses the skewness due to cosmic variance efficiently: with shape noise, none of of the COSEBIs display any significant detection of skewness anymore.

\begin{figure*}
    \centering
    \includegraphics[scale=0.57]{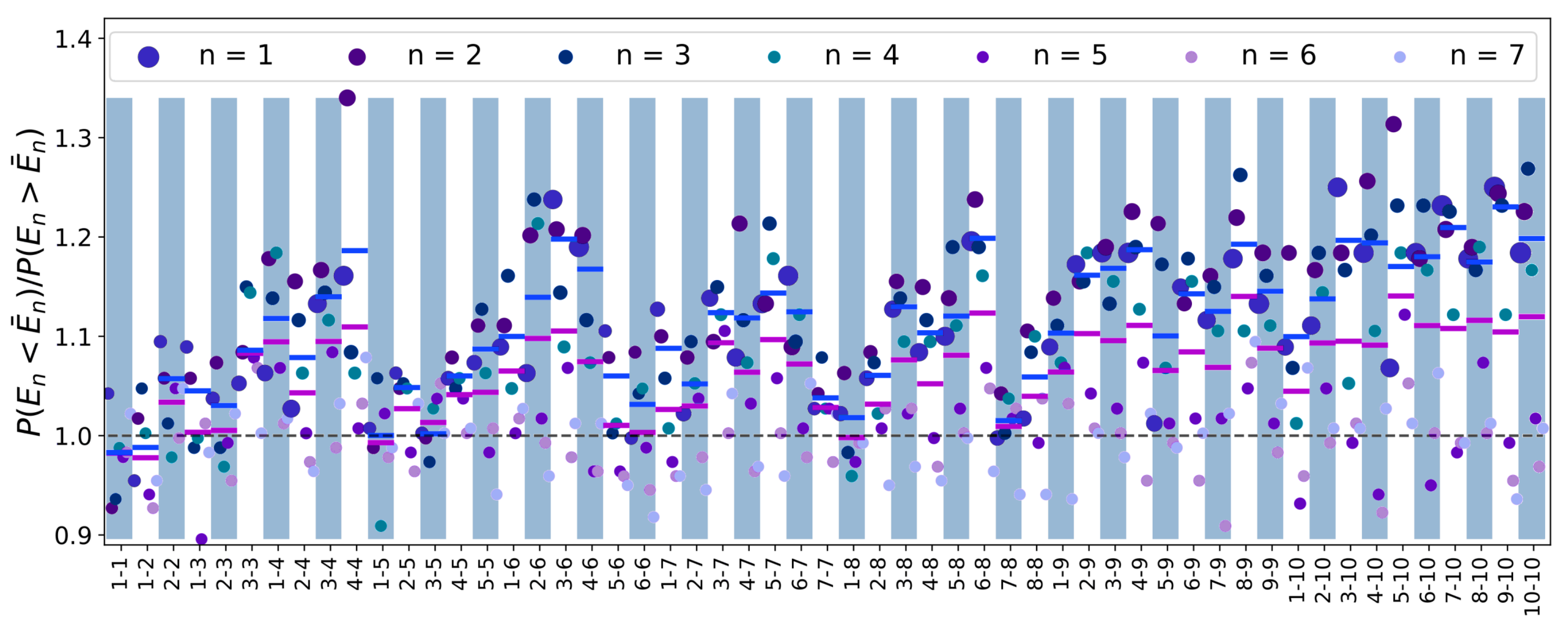}
    \caption{Excess probability of drawing a `low' weak lensing data vector in a 10-bin tomographic survey when insisting on a Gaussian likelihood. All COSEBIs here used span the angular range [1',300'], and include cosmic variance and shape noise. Plotted is the probability ratio of a COSEBI mode falling below or above its mean. For a symmetric Gaussian distribution, this ratio equals unity. The grey and white bands mark the different redshift bin combinations as indicated on the x-axis. The colourful points per band mark the probability ratio of the individual $E_n$ modes per bin. As can be seen, in particular $E_1$ often has an excess probability of 20-25\% of falling below its mean which is unexpected by a Gaussian likelihood. Moreover, the marker size of each mode is proportional to its signal-to-noise ratio (SNR), revealing that the COSEBIs contributing the highest SNR are most prominently biased low. The blue (purple) bars indicate the SNR-weighted (unweighted) average probability ratio of all modes within one redshift bin being biased low. This probability of being biased low increases whenever at least one redshift bin lies at large redshift, causing the visible sinusoidal pattern. Typical unweighted probabilities of being biased low amount to 10\%.  Weighting by the SNR, it can be seen that in 20-25\% of all cases, the signal will be systematically lower than expected from a Gaussian likelihood. Roughly a quarter of a Euclid- or LSST-sized data set will thus be affected. } %Instead of reporting by how much it will be biased low (which has been solved in SHH18, we here report how often it will be biased low
    \label{skew}
    \vspace{10pt}
\end{figure*}

\section{Excess probability of finding `low' weak lensing measurements}
\label{sec:excess}
As known from \citet{SHH18}, data points for $\xi_+(\theta)$ on angular scales of 100 arcminutes and beyond are expected to be biased low by 0.3 standard deviations in the low redshift bins of a Euclid- or LSST-like survey. This bias was shown to exacerbate with redshift. $\xi_-$ is similarly biased low, but to a lesser degree.

COSEBIs form linear combinations of $\xi_+(\theta)$ and $\xi_-(\theta)$. If the COSEBI filter is positive everywhere, any COSEBI computed from low $\xi_\pm$ will in turn be biased low itself. Only when the COSEBI filter changes sign can it be expected that biases are suppressed, see the discussion of Fig.~\ref{fig:3transcovs} in Sect.~\ref{sec:CMBcomparison}. With Euclid and LSST set up to target large angular scales, we therefore investigate the biases of COSEBIs on the [1',300'] scale, and set these biases into perspective with the signal-to-noise contributed by each COSEBI mode.

In Fig.~\ref{skew} we plot the excess probability ratio of COSEBI modes taking values below their mean. This `excess' is rated in comparison to a Gaussian. If COSEBI data points were drawn from a Gaussian distribution, they are equally likely to take values below or above their mean. 

To quantify the excess probability ratio of COSEBIs falling below the mean, we analyze all our COSEBI samples directly, without binning in a histogram. For each COSEBI mode $E_n$, we compute the fraction of samples that falls below and above the mean $\bar{E}_n$. This estimates the two probabilities $\calP(E_n < \bar{E}_n)$ and $\calP(E_n > \bar{E}_n)$. For a Gaussian, the ratio of these probabilities will equal unity, with some small scatter due to estimating the probability from 819 samples.

We define the ratio of these probabilities as
\begin{equation}
    e_n = \frac{\calP(E_n < \bar{E}_n)}{\calP(E_n > \bar{E}_n)},
\end{equation}
which is the `excess' of COSEBIs taking values below their mean.

Fig.~\ref{skew} reveals that this excess probability for the COSEBIs deviates from unity. We observe that in particular the low COSEBI modes $E_1$ to $E_4$ are at least 10\% more likely to fall below the mean than expected from a Gaussian. However, in particular for high redshifts, this excess can amount to 20-25\%, in rare cases even exceeding 30\%. 

\begin{figure*}
    \centering
    \includegraphics[scale=0.79]{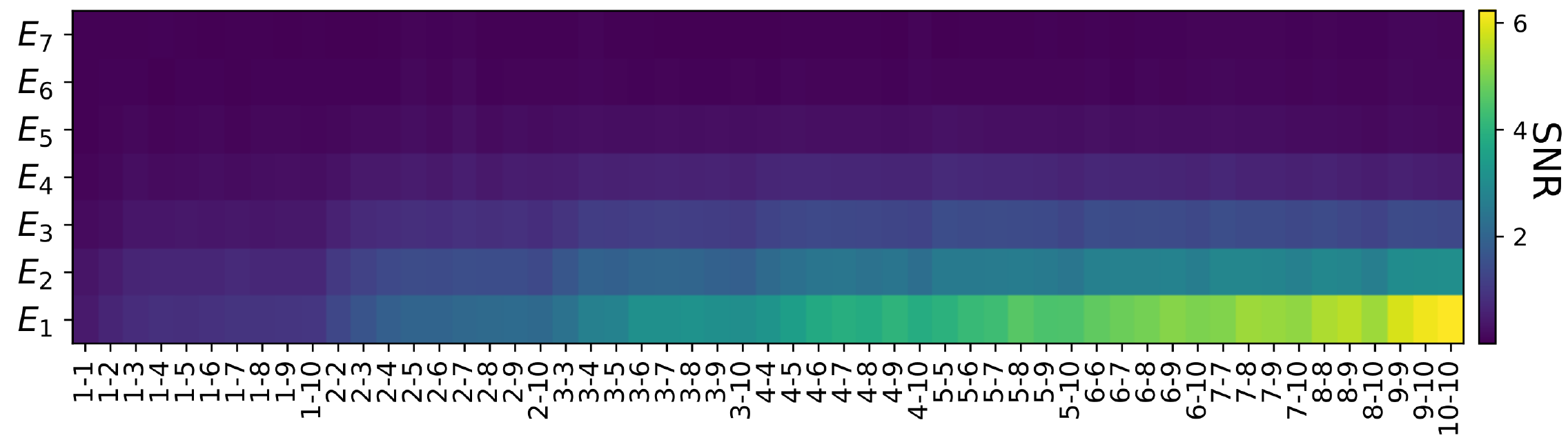}
\caption{The signal-to-noise that each $E_n$ COSEBI contributes in a 10-bin tomographic survey with shape noise. Indicated along the horizontal axis are the redshift bin combinations, along the vertical axis the index $n$ of the COSEBI E-modes. As expected from weak lensing being an integral effect, the highest signal-to-noise is carried by the highest redshift bin combinations. With increasing $n$, the COSEBI modes contribute ever less signal, motivating the truncation at $n = 7$.}
    \label{fig:SNR}
    \vspace{10pt}
\end{figure*}

Assuming all COSEBI modes contributed with equal weight to the signal in a redshift bin, we compute the arithmetic mean per redshift bin
\begin{equation}
    \bar{e}_n = \frac{1}{7}\sum_{i=1}^7 e_n.
\end{equation}
This average excess probability ratio of all COSEBIs per bin falling below their mean is indicated as purple horizontal bars in Fig.~\ref{skew}. A sinusoidal pattern of this averaged excess can be seen, which originates from the ordering of redshift bin combinations: combinations with at least one bin at high redshift are systematically more likely to yield low values for COSEBIs. In comparison to a Gaussian, Fig.~\ref{skew} indicates typically a 10\% excess of yielding a low weak lensing data vector, if all seven COSEBIs are assigned equal weights while averaging.

In reality however, the different COSEBI modes contribute a different signal-to-noise to the entire measurement. Computing an unweighted mean therefore neglects that certain COSEBIs are more influential for the final inference than others. To refine the study, we therefore proceed by computing the signal-to-noise per COSEBI, and then compute the signal-to-noise weighted expectation for low COSEBI amplitudes.

Fig.~\ref{fig:SNR} depicts the signal-to-noise ratio for the different COSEBI modes in a 10-bin tomographic survey. In Fig.~\ref{skew}, the marker size of each COSEBI mode is proportional to its signal-to-noise. We define the signal-to-noise ratio as
\begin{equation}
    {\rm SNR}(n) = \frac{\bar{E}_n}{\sigma(E_n)},
\end{equation}
where $\bar{E}_n$ is the mean of a COSEBI mode, and $\sigma(E_n)$ its standard deviation. In Fig.~\ref{fig:SNR}, the seven modes used in this study are indicated along the horizontal axes, whereas the redshift bin combinations are indicated on the vertical axis. As weak lensing integrates along the line of sight, the signal-to-noise increases with redshift. Also visible is that the information content of COSEBIs saturates with increasing $n$, implying that also within a single redshift bin combination the different COSEBI modes will have unequal impact on the total signal per bin. 

To account for this inequality of modes, we compute the weighted average excess probability
\begin{equation}
    \bar{e}_n^{\rm w} = \frac{\sum_{i = 1}^7 {\rm SNR}(n) e_n}{\sum_{i=1}^7 {\rm SNR}(n)},
\end{equation}
which weighs modes according to their signal-to-noise. These weighted excess probability ratios for the weak lensing signal per bin to fall below its mean are marked by blue bars in Fig.~\ref{skew}. As the modes with highest signal to noise are precisely those that are the most biased, the weighted average for all COSEBIs taking values below their mean is larger than its unweighted equivalent: Fig.~\ref{skew} reveals a 15-20\% excess probability ratio of all modes taking `surprisingly low' values, if regarded with a Gaussian expectation. This trend exacerbates for high redshifts.

In total, the implications for a Euclid- or LSST-like survey are clear: given Fig.~\ref{skew} there exists the choice between removing the modes of highest signal-to-noise from the data set in order to debias --- or a non-Gaussian likelihood is required for a fully bias-free inference. As both surveys are conceived to yield a precision measurement of weak lensing, the removal of the highest signal-to-noise modes does not seem to be the natural solution.

\section{Conclusions}
\label{sec:conclusions}
The sampling distribution --- and in consequence the likelihood --- of any 2-point function is a priori expected to be skewed. This has been implemented in likelihoods of the CMB, see e.g. \citet{PlanckLikelihood2018}, and while \citet{ManriqueYus1} developed a skewed core-likelihood for weak lensing, galaxy clustering and their cross-correlation, the standard in weak lensing analyses still employs a symmetric Gaussian likelihood.
Skewness in a sampling distribution implies the measurement gained most likely will systematically differ from the average measurement. As already shown in \citet{SHH18}, this affects weak lensing, and both the magnitude and direction of this bias are similar to the lowness of $\sigma_8$ as in comparison to Planck's measurements.

From a somewhat misplaced appeal to the Central Limit Theorem, one could hope that data compression reduces the bias arising from skewness. The hope therein is that large and small values of data points cancel during compression, thereby yielding a compressed data vector that scatters more tightly around its mean. This hope is somewhat misplaced, as the Central Limit Theorem implies that (potentially weighted) sums over random variables will tend towards Gaussianly distributed variables. This is still consistent with the statement that if all random variables happen to be `low', then their sum will equally be `low'. 

Cosmological data compression often takes the form of filtering, where the scalar product with the filter causes the weighted averaging. Accordingly, if the original signal is low, the filtered signal will also be low, unless the filter has specifically been constructed to cancel the low values. This argument holds for all filter-based compression techniques \citep{MOPEDI,MOPEDII,MOPEDIII,AlsingCompr} and in weak lensing also for COSEBI filters \citep{SK07,SEK10}.

COSEBIs were originally designed to split weak lensing E- and B-modes. A priori, they are thus not made to suppress biases from skewness, but could coincidentally do so. Accordingly, this paper studied whether COSEBIs suppress biases from sampling distribution skewness.

In Sect.~\ref{sec:redshift} we focussed on a Euclid- or LSST-like 10 bin tomoraphic weak lensing survey and showed that the skewness of a given COSEBI mode's sampling distribution increases with redshift. Up to a unit redshift, which comprises the majority of all galaxies observed by KiDS and DES, no significant skewness is detected, implying a COSEBI analysis of the current surveys may use a Gaussian likelihood. In contrast, roughly half of the galaxies observed with LSST or Euclid will fall into redshift ranges where the skewness of the COSEBI distribution is detected at 6$\sigma$ significance.

In Sect.~\ref{sec:CMBcomparison} we provide evidence that weak lensing observables are drawn from a much more skewed distribution than CMB observations of the same multipole. While the CMB likelihood for spherical harmonics power spectra $C_\ell$ Gaussianizes at $\ell \approx 20$, the weak lensing likelihood remains skewed until $\ell \approx 100.$

The skewness of 2-point function sampling distributions implies that if the data are analyzed with a symmetric Gaussian likelihood, then one will often find `surprisingly low' amplitudes. This arises since the Gaussian misestimates how frequently the data will take values below their mean. In Sect.~\ref{sec:excess} we showed that if a Gaussian likelihood is insisted upon for a Euclid- or LSST-like survey, then in 10\% of all cases the data points will appear to be systematically low. Particularly affected are those COSEBI modes which contribute the highest signal-to-noise, such that in the high signal-to-noise regime systematically low amplitudes are to be expected in up to 25\% of the cases. \lna{For modes affected by high shape noise, this can be understood intuitively: as shape noise adds Gaussian noise, a mode whose SNR is suppressed by shape noise is automatically a mode whose scatter is dominated by Gaussian noise. Hence these modes will show little bias, as the Gaussian shape noise will cause dominating symmetric noise. } In total, we conclude that the next generation of surveys has the choice between lowering the signal-to-noise by discarding the modes which are affected by skewness-induced biases, or to retain these modes and develop a non-Gaussian likelihood for their analysis.

The skewness here discussed is known from analyses of the cosmic microwave background; as such, it is fundamentally different from gravity-induced non-Gaussianity. The interested reader may find a further discussion of gravitationally induced non-Gaussianity in weak lensing in appendix \ref{app:gravity}.

\appendix

\section{Computation of COSEBIs}
\label{app:cosebis}
In this appendix we detail the construction of COSEBIs, following \citet{SEK10,SK07}.

COSEBIs were initially introduced in order to split E- and B-modes in a weak lensing survey. Their filters are polynomials that are linear or logarithmic in the angular scale $\theta$. As linear COSEBIs have been shown to be significantly less efficient in retaining cosmological information \citep{SEK10}, we directly used logarithmic COSEBIs in this paper. 

Unlike a correlation function that is a continuous function of angular separation $\theta$, COSEBIs are set of discrete modes, labelled by an index $n$. This index is accordingly indicated in all relevant  figures. We denote modes that only contain E-type power as $E_n$, and COSEBI B-modes by $B_n$. If a cosmology of cold dark matter (CDM), and a non-standard dark energy equation of state $w$ is to be studied, then expanding COSEBI modes until $n = 7$ is sufficient \citep{CosebiwCDM}.

To derive COSEBIs via their property of splitting E- and B-modes, we introduce an E-mode power spectrum $P^{ij}_E(\ell)$, and a B-mode power spectrum $P^{ij}_B(\ell)$, where $ij$ determines the tomographic bin combination. Weak lensing cannot induce a B-mode power spectrum, and it is likewise hoped that no systematics contribute to the E-mode power spectrum. In this case it then follows that $P_E(\ell)$ equals Eq.~(\ref{P}). 

In terms of both $P^{ij}_E(\ell)$ and a putative $P^{ij}_B(\ell)$, the weak lensing correlations functions $\xi_\pm$ are given by 
 \begin{equation}
     \xi^{ij}_{+}(\theta) = \int_0^{\infty} \frac{\mathrm{d}\ell \ell}{2\pi} \mathrm{J}_0(\theta\ell) \left[\mathrm{P}^{ij}_E(\ell)+ \mathrm{P}^{ij}_B(\ell)\right],
 \end{equation}
and
 \begin{equation}
 \xi^{ij}_{-}(\theta) = \int_0^{\infty} \frac{\mathrm{d}\ell \ell}{2\pi} \mathrm{J}_4(\theta\ell) \left[\mathrm{P}^{ij}_E(\ell)- \mathrm{P}^{ij}_B(\ell)\right],
 \end{equation}
implying they will mix E- and B-modes, if B-modes are present \citep{SK07}. To create a linear filter that splits E- from B-modes, the ansatz
%\int_{\theta_{\mathrm{min}}}^{\theta_{\mathrm{max}}}
\begin{equation}
\begin{aligned}
    E_n^{(ij)} = \frac{1}{2}\int_0^\infty \mathrm{d}\theta \theta\Big[  T_{+}^{(n)}(\theta)\xi_{+}^{(ij)}(\theta) 
     + T_{-}^{(n)}(\theta)\xi_{-}^{(ij)}(\theta)\Big],\\
    B_n^{(ij)} = \frac{1}{2}\int_0^\infty \mathrm{d}\theta \theta\Big[  T_{+}^{(n)}(\theta)\xi_{+}^{(ij)}(\theta) 
     - T_{-}^{(n)}(\theta)\xi_{-}^{(ij)}(\theta)\Big],
\end{aligned}
\label{EBfromxipm}
\end{equation}
is chosen. The filter functions $T_\pm^{(n)}(\theta)$ thus depend on the index $n$ which determines the order of the COSEBI mode.

To construct a filter which splits E- from B-modes, we transform the filters $T_\pm^{(n)}(\theta)$ to harmonic space
\begin{equation}
\begin{aligned}
    W_+(\ell) & = \int_0^\infty \mathd \theta\ \theta T^{(n)}_+(\theta)J_0(\ell \theta),\\
    W_-(\ell) & = \int_0^\infty \mathd \theta\ \theta T^{(n)}_-(\theta)J_4(\ell \theta),
    \end{aligned}
\end{equation}
as the corresponding E- and B modes in harmonic space are then simply given by
\begin{equation}
\begin{aligned}
    E_n = \int_0^\infty \frac{\mathd \ell \ \ell}{2\pi} \Big( & P_E(\ell)[W_+^{(n)}(\ell) + W_-^{(n)}(\ell)]\\
    & + P_B(\ell)[W_+^{(n)}(\ell) - W_-^{(n)}(\ell)]\Big),\\
B_n = \int_0^\infty \frac{\mathd \ell \ \ell}{2\pi} \Big( & P_E(\ell)[W_+^{(n)}(\ell) - W_-^{(n)}(\ell)]\\
    & + P_B(\ell)[W_+^{(n)}(\ell) + W_-^{(n)}(\ell)]\Big).
    \end{aligned}
    \label{EBP}
\end{equation}
This allows us to read off the defining property of a filter that splits E- and B-modes, namely the equality
\begin{equation}
    W_+^{(n)}(\ell) = W_-^{(n)}(\ell),
    \label{Weq}
\end{equation}
such that the angular brackets in Eq.~(\ref{EBP}) vanish, thereby enforcing the split.

Fortunately, infinitely many solutions for the constraint Eq.~(\ref{Weq}) exist. This enables the specification of further advantageous side constraints on the filter. COSEBIs use this freedom to impose additionally that only a finite range in $\theta$ be used. The integrals in Eq.~(\ref{EBfromxipm}) will then only run over the interval $[{\theta_{\mathrm{min}}},{\theta_{\mathrm{max}}}]$ rather than from zero to infinity. By setting $[{\theta_{\mathrm{min}}},{\theta_{\mathrm{max}}}]$ to within the bounds of the observation, we can thus enforce that only observed ranges are used. 

To exclude angular ranges, the filter $T_+^{(n)}(\theta) \equiv 0$ outside $[\theta_{\rm min},\theta_{\rm max}]$, implying the filter's zero amplitude does not let pass data from outside these angular ranges. As $T^{(n)}_+$ and $T^{(n)}_-$ are related to each other via the identity of their transforms Eq.~(\ref{Weq}), the filter $T_+^{(n)}$ must then additionally fulfil
\begin{equation}
    \int_{\theta_{\rm min}}^{\theta_{\rm max}} \mathd \theta \ \theta T_+^{(n)}(\theta) = \int_{\theta_{\rm min}}^{\theta_{\rm max}} \mathd \theta \ \theta^3 T_+^{(n)}(\theta) = 0.
    \label{zeropass}
\end{equation}
such that also the $T^{(n)}_-$ does not let pass data outside the chosen angular range.

The two constraints Eq.~(\ref{Weq}) and Eq.~(\ref{zeropass}) are of direct importance, as they enforce the E/B-mode split and the usage of observed angular ranges only. \citet{SEK10} furthermore impose the constraint that the sequence of filters be orthonormal, which is convenient, but not necessary. Importantly however, \citet{SEK10} show that cosmological information is better retained if the T-filters are logarithmic in $\theta$, thus motivating the last transformation
\begin{equation}
    T_+^{(n)}(\theta) = t_+^{(n)}(z), \ \ \  z = \ln(\theta/\theta_{\rm min}).
    \label{lastEq}
\end{equation}
At this stage, Eq.~(\ref{Weq}) till Eq.~(\ref{lastEq}) fully specified the scientific aims to be met with COSEBIs. The constraints following therefrom can now be met, implying the COSEBI-filters can now be explicitly constructed.
To do so, we follow Eq.~(28) -- Eq.~(39) of \citet{SEK10}, such that we compute our polynomial COSEBIs via their roots. This yields the filters shown in Fig.~\ref{Wfilters} and Fig.~\ref{Tfilters} which are used to gain the results on skewness in this paper.

\section{Discussion of gravitational skewness}
\label{app:gravity}
This paper has focused on a \emph{quantification} of skewness in the sampling distribution and likelihood of 2-point functions derived from weak lensing observations. This quantification is required to understand biases during parameter inference. Nonetheless, it transpired that much interest exists to also understand the role of gravity in this skewness. 

This interest is in particular caused by the excess skewness of weak lensing distributions (until $\ell \approx 100$), in comparison to the likelihood of the CMB power spectrum, which is only noticeably skewed until $\ell \approx 20$. While the impact of shape noise and cosmic variance on the skewness has already been factored apart in \citet{SHH18}, gravity is often put forward as a potential additional cause, and we shall thus discuss it here.

We here observed the trend that excess skewness increases with redshift, which is also consistent with previous studies on gravitationally caused non-Gaussianity: \cite{ClaudeGravityBispect} show that the gravitational bispectrum is dominated by triangles of the squeezed and elongated type, consistent with our observed redshift trend, and \cite{WLBispectByGravity} show that the weak lensing bispectrum is also dominated by squeezed configurations. The amplitude of the weak lensing bispectrum equally grows with redshift, again consistent with the findings of this paper. We limit ourselves to describing these as `consistencies' and refrain from attributing the excess skewness to gravity alone.

\section*{Acknowledgements}
ES is funded by the Oort fellowship programme of Leiden Observatory. This research has profited from academic discussions with Koen Kuijken, Henk Hoekstra and Tom Kitching. We thank Henk Hoekstra, Simon Portegies Zwart and Leiden's IT group for sharing their clusters andwelcoming this paper's greedy CPU demands. We thank Joachim Harnois-Deraps for making public the SLICS mock data, which can be found at http://slics.roe.ac.uk/.

\bibliographystyle{mnras}%For reasons not understood, the OpenJournal StyleFile needs to know that we want to have MNRAS-formatted citation style
\bibliography{TDist.bib}

\end{document}